\documentclass{article}
\usepackage{spconf,amsmath,graphicx,hyperref}
\usepackage{dblfloatfix}
\usepackage{xcolor}
\usepackage{comment}
\usepackage{pifont}

\usepackage{amsmath, amssymb, amsfonts, mathtools}

\usepackage{newunicodechar}
\usepackage[ruled,vlined,linesnumbered]{algorithm2e}
\usepackage{float} 
\usepackage{booktabs} 
\usepackage{amssymb}  
\usepackage{array}    

\usepackage{booktabs,makecell,array,pifont}

\usepackage{amsmath,amssymb,amsfonts,mathtools}
\usepackage{makecell}
\usepackage{multirow}
\usepackage{booktabs}
\usepackage{hyperref}
\usepackage{array}
\usepackage{times}     

\newcolumntype{L}[1]{>{\raggedright\arraybackslash}p{#1}}

\usepackage{etoolbox}

\usepackage{bibspacing}

\title{\vspace{-3em}MOSA: Mixtures of Simple Adapters Outperform Monolithic Approaches in LLM-based Multilingual ASR\vspace{-0.8em}}

\name{\vspace{-1em}Junjie Li$^{1}$, Jing Peng$^{1}$, Yangui Fang$^2$, Shuai Wang$^{3}$, Kai Yu$^{1}$$^{\dagger}$\thanks {$^{\dagger}$Corresponding author.}}
\address{
\small
$^1$X-LANCE Lab, School of Computer Science, Shanghai Jiao Tong University, Shanghai, China\\
\small
$^1$MoE Key Lab of Artificial Intelligence, $^1$Jiangsu Key Lab of Language Computing\\
\small
$^2$School of Electronic Information and Communications, Huazhong University of Science and Technology, China \\ 
\small
$^3$School of Intelligence Science and Technology, Nanjing University, China \\
\small
\{junjieli, jing.peng, kai.yu\}@sjtu.edu.cn~~~fangyg@hust.edu.cn~~~shuaiwang@nju.edu.cn~~~ 
\vspace{-1.3em}}

\begin{document}
\ninept
\maketitle

\begin{abstract} 
    LLM-based ASR overcomes multilingual data scarcity by projecting speech representations into the LLM space to leverage its robust semantic and reasoning capabilities.
    However, while previous approaches typically enhance performance by scaling data or model parameters, a single projector often struggles to effectively align representations across different languages. 
    In this work, we propose an MoE-based projector named MOSA (Mixture of Simple Adapters). By aggregating multiple simple adapters, this architecture enables different experts to specialize in learning either language-shared or language-specific knowledge. This approach not only mitigates parameter interference between languages but also facilitates positive transfer from high-resource to low-resource languages, effectively alleviating data scarcity issues. 
    Experimental results demonstrate that MOSA-Base achieves a 15.4\% relative reduction in average WER compared to the Ideal-LLM Base, consistently outperforming it across all languages. Notably, MOSA achieves a 13.3\% WER reduction over the Ideal-LLM Base while utilizing only 60\% of its parameters. These findings highlight MOSA's superior parameter efficiency and robustness against data imbalance, suggesting that a mixture of simple adapters is more suitable for multilingual LLM-based ASR than complex single-adapter designs.
\end{abstract}


\vspace{-0.6em}
\section{Introduction}
\vspace{-0.6em}

End-to-end multilingual automatic speech recognition (ASR) requires a single model to accurately transcribe speech from multiple languages into corresponding text. This technology plays a vital role in facilitating cross-lingual communication, such as real-time translation, and in preserving linguistic diversity. However, it also faces significant challenges, including data scarcity and multilingual parameter interference.
In recent years, large language models (LLMs) have demonstrated remarkable capabilities across various domains~\cite{radford2019language, brown2020language, larenz2023overcoming, chowdhery2023palm, anil2023palm, touvron2023llama, peng2024survey}. 
The strong performance of LLMs has sparked growing interest in a new ASR paradigm: aligning the speech input space with LLM input spaces through an projector module. This alignment enables direct utilization of the LLMs’ extensive commonsense knowledge and contextual understanding learned from large-scale text corpora.

\begin{figure*}[htbp]
    \centering
    \includegraphics[width=0.6\textwidth]{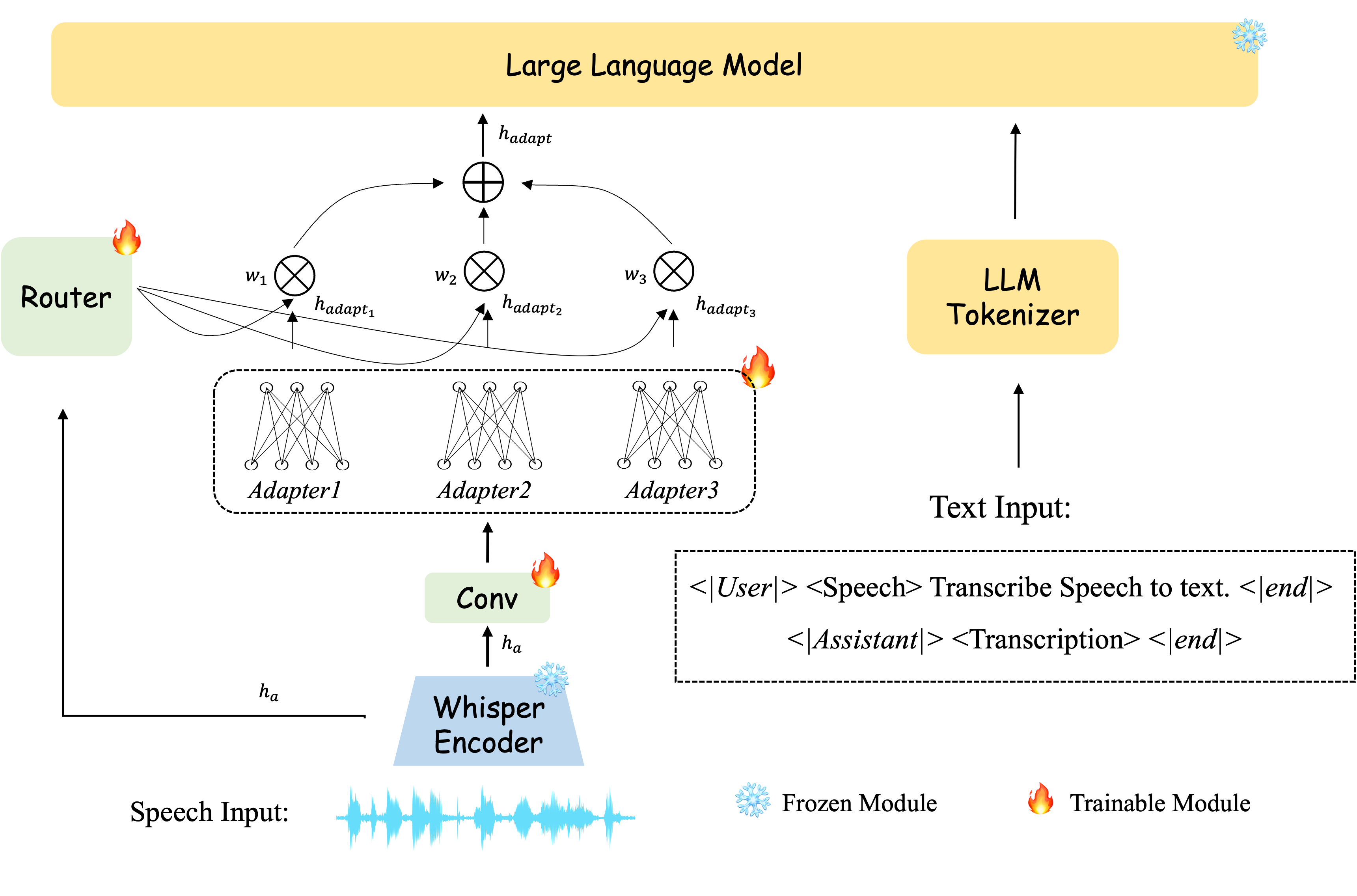}
    \vspace{-1.4em}
    \caption{Model Architecture. The speech encoder extracts features from speech. Adapters map these features into the LLM input space, guided by a Router that dynamically weights their outputs. The LLM then performs speech recognition based on the aligned representation and instruction.}
    \label{fig:model}
    \vspace{-1.2em}
\end{figure*} 

The paradigm of aligning speech encoders with LLM input spaces via projectors has been extensively explored in ASR. For instance, SLAM-ASR~\cite{ma2024embarrassingly} achieves state-of-the-art performance on the English LibriSpeech~\cite{panayotov2015librispeech} dataset by training a simple linear projector layer. \cite{geng2024unveiling} systematically compares the effects of various combinations of speech encoders, LLMs, and projector architectures on Chinese ASR performance. FireRedASR~\cite{xu2025fireredasr} employs a more powerful speech encoder to enhance Mandarin speech recognition performance.
In multilingual ASR, \cite{fathullah2024prompting} investigates the impact of frame stacking to reduce the sequence length fed into LLMs, exploring how different frame rates influence recognition accuracy. 
Qwen-Audio~\cite{chu2023qwen} leverages a carefully crafted multi-task training framework to enable multilingual speech recognition, while Qwen2-Audio~\cite{chu2024qwen2} further enhances understanding by replacing task-specific labels with natural language descriptions, thus better utilizing the reasoning capabilities of LLMs.
Seed-ASR~\cite{bai2024seed} introduces a multi-stage training strategy inspired by LLMs, which enables both multilingual recognition and improved context-guided understanding.
Ideal-LLM~\cite{xue2024ideal} proposes a dual-encoder architecture that incorporates a language classifier to assign different speech encoder weights for different languages, thereby improving multilingual adaptability.

While previous work has improved multilingual ASR performance by enhancing the speech encoder or even fine-tuning LLMs~\cite{xu2025fireredasr, chu2023qwen, chu2024qwen2, bai2024seed}, relatively little attention has been paid to the design of projector modules~\cite{xue2024ideal, mu2024hdmole}. On one hand, even if these projectors are complex and include many transformer layers, a single projector may still fail to effectively align speech representations from all languages with the LLM input space. On the other hand, most prior approaches focus on increasing the amount of training data~\cite{xu2025fireredasr, chu2023qwen, chu2024qwen2, bai2024seed}, without fully exploiting the potential for knowledge sharing across languages, such as syntactic or phonetic similarities~\cite{yu2023master}, which could benefit low-resource languages by transferring knowledge from high-resource ones.
Meanwhile, each expert in the Mixture-of-Experts (MoE)~\cite{shazeer2017outrageously, fedus2022review, zoph2022st, xie2023moec, song2024u2++} model specializes in different domains, making it a highly effective architectural design for multilingual and multi-domain settings.
HDMoLE~\cite{mu2024hdmole} addresses multi-accent ASR using a hierarchical MoE~\cite{shazeer2017outrageously} framework, where the LoRA~\cite{hu2021lora, lora_whisper} expert weights assigned by a global pre-trained accent recognition model and local models are used to handle different accents.

In this work, we posit that cross-lingual scenarios involve both shared and language-specific knowledge. A single, monolithic projector often struggles to effectively capture and map these diverse knowledge types.
To address this limitation, we propose MOSA (Mixture of Simple Adapters) within the projector, a novel approach leveraging the MoE mechanism. Each expert adapter specializes in learning either shared or language-specific features, while a router dynamically predicts expert weights for each speech input, enabling a dynamic mixture of experts.
Prioritizing simplicity, we utilize a single encoder, where each adapter consists of only two linear layers with ReLU~\cite{agarap2018deep} activation. Departing from prior methods that rely on complex loss objectives, we exclusively optimize the cross-entropy loss on transcriptions. 
Experimental results demonstrate that MOSA-Base achieves a 15.4\% relative reduction in average Word Error Rate (WER) compared to the Ideal-LLM Base, consistently outperforming it across all languages. Notably, MOSA surpasses the Ideal-LLM Base even with only 60\% of the parameter count. Furthermore, MOSA exhibits superior robustness to data imbalance. Ablation studies confirm that MOSA effectively manages multilingual alignment to mitigate parameter interference, while facilitating positive transfer from high-resource to low-resource languages to mitigate data scarcity issues.
These findings suggest that for LLM-based ASR, a mixture of simple adapters constitutes a more effective design strategy than a single, complex projector.

\vspace{-0.6em}
\section{Method}
\vspace{-0.6em}

\subsection{Mixture of Experts}
\label{Mixture of Experts}

The Mixture-of-Experts (MoE)~\cite{shazeer2017outrageously, fedus2022review, zoph2022st, xie2023moec, song2024u2++} architecture consists of two main components: a set of $N$ experts $\{E_i\}_{i=1}^N$, where each expert is expected to learn specialized or shared knowledge, and a gating network or router $G$, which is responsible for dynamically determining the contribution or weight of each expert based on the input. Given an input $x$, the weights for the experts are computed as:

\begin{equation}
w = \text{SoftMax}(G(x))
\end{equation}

The final output is then obtained by computing a weighted sum of the expert outputs. In this work, we use softmax gating instead of a sparse MoE based on top-$k$ selection:

\begin{equation}
y = \sum_{i=1}^N w_i E_i(x)
\end{equation}
where $w_i$ denotes the weight assigned to the $i$-th expert $E_i$.

\vspace{-0.2em}
\subsection{Model Architecture}

As shown in Figure \ref{fig:model}, MOSA consists of a speech encoder, a pre-trained large language model, and a projector module incorporating a MoE mechanism.

\begin{table*}[ht]
\centering
\caption{WER(\%) results on Multilingual LibriSpeech of different models. Lower is better. Ideal-LLM and MOSA share the same LLM.}
\label{tab: main results}
\begin{tabular*}{\textwidth}{@{\hspace{6mm}\extracolsep{\fill}} l c c c c c c c c c c @{\hspace{6mm}}}
\toprule
\multicolumn{1}{c}{Model}           & Trainable Params (B) & en   & de   & nl    & fr   & es   & it    & pt    & pl   & Avg  \\ \midrule
LLaMA with ASR\cite{fathullah2024prompting}  &            0.240          &   6.20   &   6.70   &   11.30    &  5.50    &   5.20   &   10.80    &    16.20   &   15.90   &  9.73    \\
Ideal-LLM Base\cite{xue2024ideal}  & 0.172               & 7.44 & 8.25 & 12.47 & 6.71 & 5.47 & 11.84 & 10.87 & 9.38 & 9.05 \\
Ideal-LLM Large & 0.303              & 6.15 & 7.12 & 11.23 & 5.40  & 4.26 & 9.93  & 12.41 & \textbf{6.02} & 7.81 \\ \midrule
MOSA-Base       & 0.155               & 5.91 & \textbf{6.30}  & \textbf{11.05} & 5.27 & 4.47 & 9.99  & 9.68  & 8.61 & 7.66 \\
MOSA-Large      &           0.287           & \textbf{5.25}      &  6.33    &  11.50    &  \textbf{5.05}    &   \textbf{4.17}   &     \textbf{9.84}  &   \textbf{9.03}    &  8.84    & \textbf{7.50} \\ \bottomrule
\end{tabular*}
\vspace{-1.2em}
\end{table*}

\textbf{Speech Encoder.}
For the speech encoder, we use the encoder component of Whisper-large-v3\footnote{\url{https://huggingface.co/openai/whisper-large-v3}}. Whisper~\cite{radford2023robust} is a multi-task, multilingual model based on an encoder-decoder architecture, which is trained on a large and diverse multilingual dataset and provides extensive coverage of various languages and acoustic patterns.

\textbf{Projector.}
The Projector module aligns the speech input space with the LLM text input space. We apply the MoE mechanism introduced in Section~\ref{Mixture of Experts} to this module. Specifically, the Projector consists of a Router and $N$ Adapters, where each Adapter is capable of learning shared or language-specific information across different languages.
Given a speech input $x$, the speech encoder $E_a$ first extracts a hidden representation $h_a$, which is then fed into both the Router and the Adapter modules. The Router computes a weight vector $w$ based on $h_a$ by pooling over the time axis, representing the contribution of each Adapter.
Before being passed to the Adapters, $h_a$ is downsampled through two convolutional layers with a stride of 2, which is then transformed by different Adapters, yielding $h_{\text{adapt}_i}$. This operation allows each element in the $h_{\text{conv}}$ sequence to incorporate information from neighboring frames. Finally, the aligned representation $h_{\text{adapt}}$ is obtained by taking a weighted sum of $h_{\text{adapt}_i}$ using the weight $w$.
The process is formulated as follows:
\begin{equation} 
\begin{aligned} 
h_a &= E_a(x) \\ 
w &= \text{Pool}_{\text{Avg}}(\text{Router}(h_a)) \\
h_{\text{adapt}_i} &= \text{Adapter}_i(\text{Conv}(h_a)) \\
h_{\text{adapt}} &= \sum_{i=1}^N w_i h_{\text{adapt}_i} 
\end{aligned} 
\end{equation}

The architectures of Adapter and Router only consist of simple linear layers with ReLU activation.
Compared to Transformer~\cite{vaswani2017attention} or Q-former~\cite{kim2024towards} based adapters commonly used in previous work, this is a much simpler design.
Our key idea is that multilingual representations are better supported by an ensemble of simple, complementary models that share representational responsibilities, rather than by a single, highly complex model, which often struggles to adequately capture the diversity and fine-grained structure of multilingual data.

\textbf{Text Decoder.}
For the pretrained large language model, following the Ideal-LLM~\cite{xue2024ideal} framework, we adopt the Phi-3-mini-4k-instruct~\cite{abdin2024phi} model\footnote{\url{https://huggingface.co/microsoft/Phi-3-mini-4k-instruct}}, which contains 3.8 billion parameters. 
The aligned speech representation $h_{\text{adapt}}$ is embedded into the user input section, forming the following structure: \texttt{<|user|> <$h_{\text{adapt}}$> Instruction <|end|> <|assistant|> Transcription <|end|>}.
The text labels and instruction prompt are tokenized using the LLM tokenizer and converted into input embeddings.
Finally, the LLM processes the aligned representation to understand the audio input and conditioned on the instruction, performs speech recognition. 
The cross-entropy loss is applied exclusively to the transcription tokens, ensuring that the model focuses on accurate speech recognition.

\vspace{-0.6em}
\section{Experiments}
\vspace{-0.6em}

\subsection{Datasets}

We adopt the Multilingual LibriSpeech (MLS)~\cite{pratap2020mls} dataset as our multilingual ASR benchmark. This dataset contains approximately 50,000 hours of speech across eight languages: English (en), German (de), Dutch (nl), French (fr), Spanish (es), Italian (it), Portuguese (pt), and Polish (pl).
Despite the large total volume, the data distribution is heavily skewed: English dominates with over 44,600 hours, while the remaining languages range from only 100 to 2,000 hours.
Before computing the WER, we apply Whisper’s multilingual normalization\footnote{\url{https://github.com/openai/whisper/blob/main/whisper/normalizers/basic.py}}, which replaces any non-standard markers, symbols, or punctuation with spaces while preserving diacritics.

\vspace{-0.2em}
\subsection{Implementation Details}

The output of the Whisper encoder, denoted as $h_a$, corresponds to 20 ms segments of the original audio, which is further downsampled to obtain $h_{\text{conv}}$ with a temporal resolution of 12.5 Hz. 
SpecAugment~\cite{park2019specaugment} is applied to speech input.
The instruction prompt is: \textit{Transcribe speech to text}. Both the speech encoder and the LLM are kept frozen during training.

To enable a fair comparison with Ideal-LLM~\cite{xue2024ideal}, we implement both \texttt{Base} and \texttt{Large} variants of our model. 
The Adapter is consistent across both variants: two linear layers with a ReLU activation in between, projecting the hidden dimension from 3072 to 4096 and back to 3072.
The main difference between the two variants lies in the number of Adapters used: the \texttt{Base} version includes 4 Adapters, whereas the \texttt{Large} version uses 8.
For the Router module, the \texttt{Base} version is composed of two linear layers with a ReLU activation in between, mapping from 1280 to 512 and finally to 4 output dimensions. In contrast, the \texttt{Large} version employs a deeper architecture with five linear layers and intermediate ReLU activations, with dimensions: 1280 $\rightarrow$ 2560 $\rightarrow$ 5120 $\rightarrow$ 2560 $\rightarrow$ 1280 $\rightarrow$ 8.

For training, we use the AdamW~\cite{loshchilov2017decoupled} optimizer with $\beta_1 = 0.9$ and $\beta_2 = 0.999$. The maximum learning rate is set to $5\mathrm{e}{-4}$ for the \texttt{Base} version and $2\mathrm{e}{-4}$ for the \texttt{Large} version. We apply 2000 warmup steps followed by a linear decay learning rate scheduler. 
Training is conducted on 8 GPUs. Each device uses a batch size of 8 with gradient accumulation steps set to 16, processing approximately 1280 seconds of audio per device. 
The \texttt{Base} version is trained for 11k steps, while the \texttt{Large} version is trained for 36k steps.
Following Ideal-LLM, both the \texttt{Base} and \texttt{Large} versions use the complete training data for the other seven languages. However, the \texttt{Base} version only uses 10k hours of English data, while the \texttt{Large} version uses the full 44k hours of English data.
Finally, due to the extreme imbalance among languages in the dataset, we follow~\cite{conneau2020unsupervised} to construct multilingual batches.

\begin{table*}[ht]
\centering
\caption{WER(\%) results on Multilingual Librispeech of different number of adapters. Lower is better. $1^*$ denotes results obtained by separately fine-tuning for each individual language, on top of the model of 1 adapter.}
\label{tab: abla1}

\begin{tabular*}{\textwidth}{@{\hspace{6mm}\extracolsep{\fill}} c c c c c c c c c c c @{\hspace{6mm}}}
\toprule
\multicolumn{1}{c}{\#Adapter}                & Trainable Params (B) & en   & de   & nl    & fr   & es   & it    & pt   & pl   & Avg  \\ \midrule
1 & 0.079 & 6.32 & 6.82 & 11.86 & 5.46 & 4.55 & 10.64 & 9.38 & 10.51 & 8.19 \\
$1^*$ & 0.079 & 6.17 & 7.00 & 11.91 & 5.47 & 4.54 & 10.34 & 10.16 & 10.53 & 8.27 \\
2 & 0.104               & 6.12 & 6.46 & 11.46 & 5.45 & 4.53 & 10.55 & 8.95 & 9.26 & 7.85 \\
3 & 0.130               & 6.32 & 6.36 & 11.32 & 5.29 & 4.64 & 9.99  & 9.04 & 9.01 & 7.75 \\
4 & 0.155               & 5.91 & 6.30  & 11.05 & 5.27 & 4.47 & 9.99  & 9.68 & 8.61 & 7.66 \\
5 & 0.180               & 5.97 & 6.23 & 11.14 & 5.10  & 4.43 & 9.94  & 9.95 & 9.16 & 7.74 \\ \bottomrule
\end{tabular*}
\vspace{-1.2em}

\end{table*}

\vspace{-0.2em}
\subsection{Main Results}

Table~\ref{tab: main results} presents the WER (\%) results on the test sets of 8 languages from the MLS dataset. Among the baselines\footnote{These results are taken from the Ideal-LLM paper.}, Ideal-LLM~\cite{xue2024ideal} adapts to multilingual settings by assigning weights to a dual-encoder architecture using a language weight, allowing different encoders to contribute differently depending on the language of the input. LLaMA with ASR~\cite{fathullah2024prompting} integrates a CTC-trained encoder with LLaMA for speech recognition and explores several factors affecting performance, such as encoder size and stride, as well as LoRA-based fine-tuning of LLaMA.
MOSA-Base achieves an average WER reduction of 21.3\% and 15.4\% compared to LLaMA with ASR and Ideal-LLM Base, respectively. It also outperforms both baselines across all 8 languages. Furthermore, MOSA requires fewer training parameters, making it a more efficient architecture that achieves state-of-the-art results.
These results suggest that a single, large adapter may struggle to align multilingual representations with the LLM space. In contrast, combining multiple lightweight adapters, each responsible for either language-specific or shared knowledge, enables a simpler architecture to achieve superior performance.
For MOSA-Large, the average WER is further improved. MOSA demonstrates a better ability to handle data imbalance. On the two languages with the least data, Polish and Portuguese.
While MOSA-Large exhibits only marginal degradation in Polish compared to MOSA-Base, it demonstrates substantial improvements in Portuguese. Conversely, although Ideal-LLM Large yields better results in Polish, it suffers from severe degradation in Portuguese. Consequently, MOSA-Large achieves overall performance that is superior or comparable to that of Ideal-LLM Large.

\vspace{-0.2em}
\subsection{Ablation Study}

Since the only difference in training data between the \texttt{Base} and \texttt{Large} versions lies in the English portion, the impact on other languages is minimal. Furthermore, the Large version requires significantly more computational resources, so we conduct the ablation studies on the \texttt{Base} version.

\textbf{A. Impact of the Number of Adapters.} 
Table \ref{tab: abla1} presents the impact of varying the number of adapters on ASR performance. Regardless of whether 2, 3, 4, or 5 adapters are used, the average WER and monolingual WER consistently outperform the Ideal-LLM Base model. In other words, even with just two adapters (60\% training parameters of Ideal-LLM Base), the performance surpasses Ideal-LLM Base by 13.3\%. As the number of adapters increases, the average WER gradually decreases, and the performance across most individual languages also improves. However, using five adapters does not yield further improvement, likely due to the model becoming too large relative to the available training data.
Furthermore, we investigated the performance of using a single adapter without incorporating a router. The results indicate that, across multiple languages, the single-adapter setup consistently underperforms compared to the multi-adapter configuration. This is particularly evident in low-resource languages such as Polish. The relatively strong performance on Portuguese may be attributed to the adapter being optimized for that specific language at checkpointing.
To further examine this, we fine-tuned the single-adapter model individually on the data of each language, resulting in eight distinct systems (denoted as $1^*$). However, the improvements observed were still marginal across all languages. These findings suggest that multiple adapters are capable of capturing language-specific or shared knowledge, thereby facilitating positive transfer from high-resource to low-resource languages.

\begin{figure}[h]
  \centering
  \includegraphics[width=0.8\linewidth]{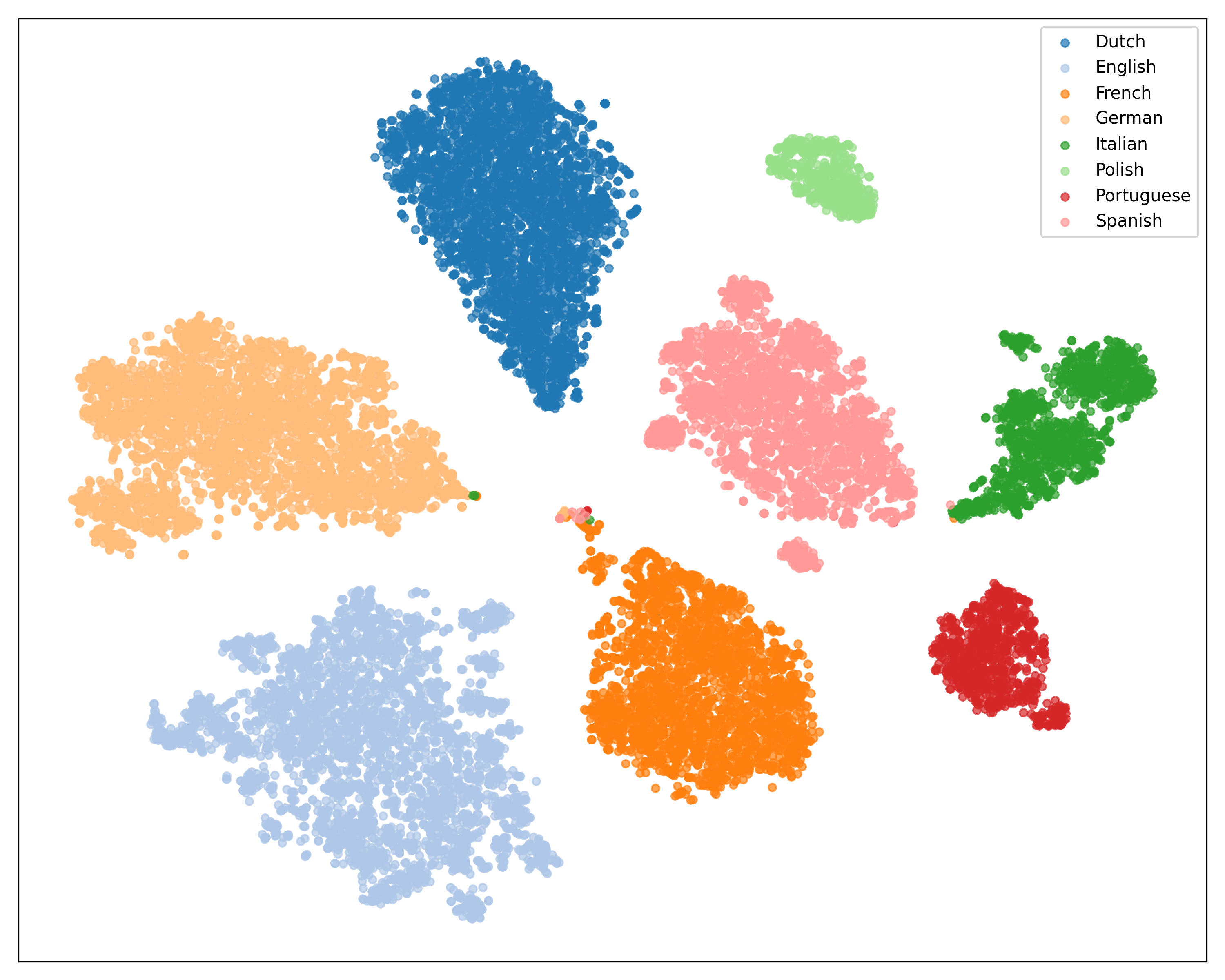}
  \vspace{-1.2em}
  \caption{T-SNE visualization of aligned speech embeddings.}
  \label{fig:abla1}
  \vspace{-0.6em}
\end{figure}

\textbf{B. Analysis of Speech Embeddings Across Languages.}
We conducted a distributional analysis of the speech embeddings generated by the projector of MOSA-Base. Specifically, we applied t-SNE~\cite{van2008visualizing} visualization on the test set of the MLS dataset, as shown in Figure~\ref{fig:abla1}. The results reveal that the embeddings from all eight languages in MLS form clearly separable clusters, with only a few outlier utterances. This indicates that the model is capable of treating each language similarly to a monolingual system, thereby mitigating multilingual parameter interference. These findings further demonstrate that mixing multiple adapters can better accommodate multilingual settings.

\begin{figure}[htbp]
  \centering
  \includegraphics[width=1.0\linewidth ]{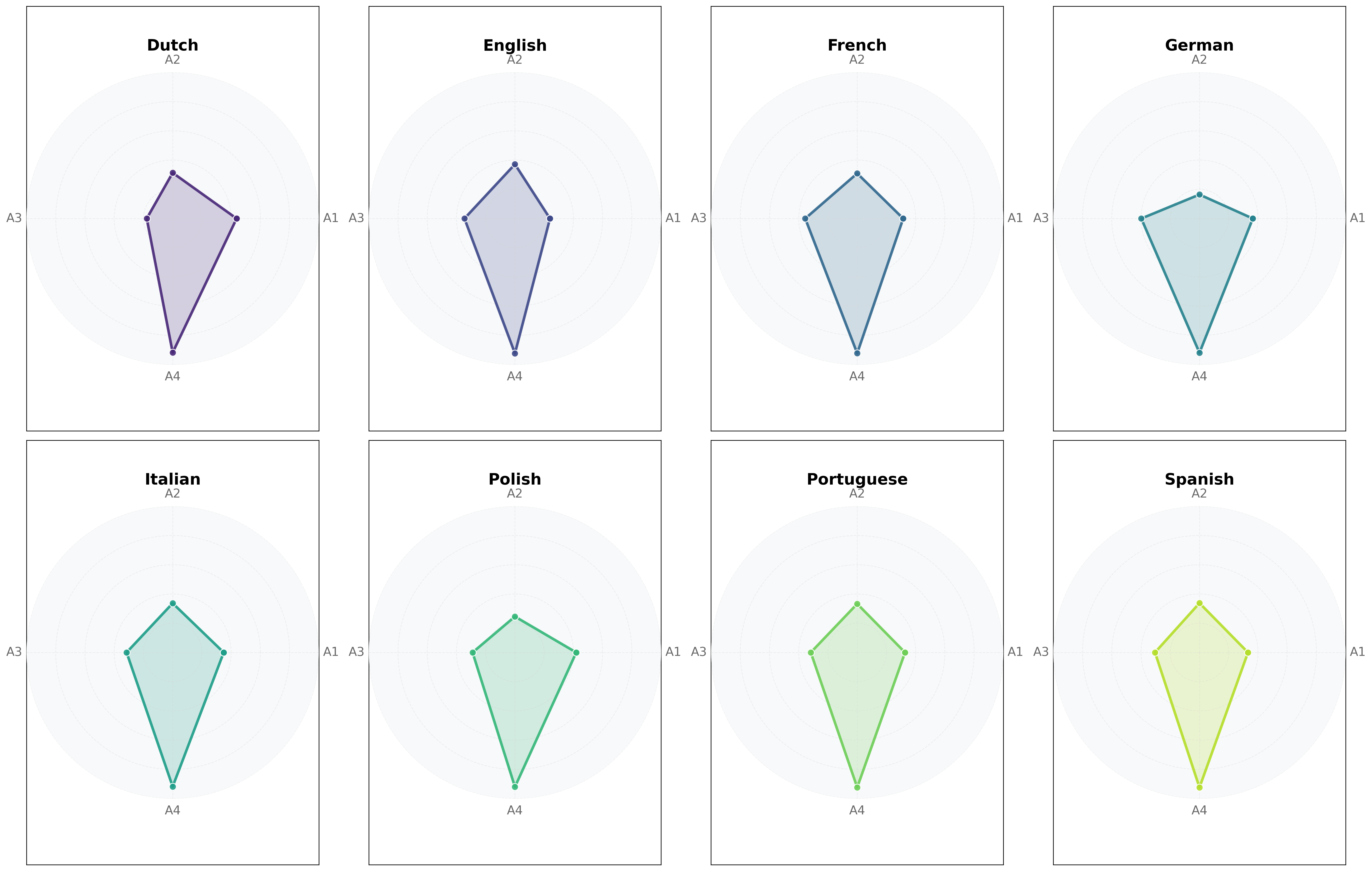}
  \vspace{-1.2em}
  \caption{Adapter weight distribution across languages.}
  \label{fig:abla2}
  \vspace{-0.6em}
\end{figure}

\textbf{C. Analysis of Adapter Weights Across Languages.}
We further analyzed the adapter weight distribution of MOSA-Base across different languages, reflecting each language's preference for specific adapters. The results, illustrated in Figure~\ref{fig:abla2}, show that all languages heavily utilize Adapter $4$, while the remaining three adapters receive varying weights depending on the language. This suggests that there exists shared knowledge across languages, which is captured by the commonly used adapter.
Meanwhile, the other adapters are likely responsible for learning language-specific knowledge, functioning as a mechanism to adjust or refine the shared representations.

\vspace{-0.6em}
\section{Conclusion}
\vspace{-0.6em}

In this work, we observe that a single, complex projector struggles to effectively align representations across multiple languages. To address this, we propose MOSA within the projector, which combines multiple lightweight adapters. This design enables different experts to capture either language-specific or shared knowledge, thereby mitigating multilingual parameter interference and facilitating positive transfer from high-resource to low-resource languages.
Experimental results indicate that MOSA-Base achieves a 15.4\% relative improvement in average WER over Ideal-LLM Base, consistently outperforming it across all evaluated languages. Notably, MOSA maintains superior performance even when trained with only 60\% of the parameters used by Ideal-LLM Base, demonstrating its exceptional parameter efficiency and robustness to data imbalance.

\vspace{-0.6em}
\section{Acknowledgment}
\vspace{-0.6em}
This work was supported by the Yangtze River Delta Science and Technology Innovation Community Joint Research Project (2024CSJGG1100).

\clearpage
\bibliographystyle{IEEEbib}
\bibliography{strings,refs}

@article{brown2020language,
  title={Language models are few-shot learners},
  author={Brown, Tom and Mann, Benjamin and Ryder, Nick and Subbiah, Melanie and Kaplan, Jared D and Dhariwal, Prafulla and Neelakantan, Arvind and Shyam, Pranav and Sastry, Girish and Askell, Amanda and others},
  journal={Advances in neural information processing systems},
  volume={33},
  pages={1877--1901},
  year={2020}
}

@article{radford2019language,
  title={Language models are unsupervised multitask learners},
  author={Radford, Alec and Wu, Jeffrey and Child, Rewon and Luan, David and Amodei, Dario and Sutskever, Ilya and others},
  journal={OpenAI blog},
  volume={1},
  number={8},
  pages={9},
  year={2019}
}

@article{larenz2023overcoming,
  title={On Overcoming the Alleged Alienation between Christianity and Physics},
  author={Larenz, Rudolf},
  year={2023}
}

@article{chowdhery2023palm,
  title={Palm: Scaling language modeling with pathways},
  author={Chowdhery, Aakanksha and Narang, Sharan and Devlin, Jacob and Bosma, Maarten and Mishra, Gaurav and Roberts, Adam and Barham, Paul and Chung, Hyung Won and Sutton, Charles and Gehrmann, Sebastian and others},
  journal={Journal of Machine Learning Research},
  volume={24},
  number={240},
  pages={1--113},
  year={2023}
}

@article{anil2023palm,
  title={Palm 2 technical report},
  author={Anil, Rohan and Dai, Andrew M and Firat, Orhan and Johnson, Melvin and Lepikhin, Dmitry and Passos, Alexandre and Shakeri, Siamak and Taropa, Emanuel and Bailey, Paige and Chen, Zhifeng and others},
  journal={arXiv preprint arXiv:2305.10403},
  year={2023}
}

@article{touvron2023llama,
  title={Llama: Open and efficient foundation language models},
  author={Touvron, Hugo and Lavril, Thibaut and Izacard, Gautier and Martinet, Xavier and Lachaux, Marie-Anne and Lacroix, Timoth{\'e}e and Rozi{\`e}re, Baptiste and Goyal, Naman and Hambro, Eric and Azhar, Faisal and others},
  journal={arXiv preprint arXiv:2302.13971},
  year={2023}
}

@inproceedings{radford2023robust,
  title={Robust speech recognition via large-scale weak supervision},
  author={Radford, Alec and Kim, Jong Wook and Xu, Tao and Brockman, Greg and McLeavey, Christine and Sutskever, Ilya},
  booktitle={International conference on machine learning},
  pages={28492--28518},
  year={2023},
  organization={PMLR}
}

@article{ma2024embarrassingly,
  title={An Embarrassingly Simple Approach for LLM with Strong ASR Capacity},
  author={Ma, Ziyang and Yang, Guanrou and Yang, Yifan and Gao, Zhifu and Wang, Jiaming and Du, Zhihao and Yu, Fan and Chen, Qian and Zheng, Siqi and Zhang, Shiliang and others},
  journal={arXiv preprint arXiv:2402.08846},
  year={2024}
}

@inproceedings{lora_whisper,
  title     = {{LoRA-Whisper}: Parameter-Efficient and Extensible Multilingual {ASR}},
  author    = {Zheshu Song and Jianheng Zhuo and Yifan Yang and Ziyang Ma and Shixiong Zhang and Xie Chen},
  year      = {2024},
  booktitle = {Proc. Interspeech},
  address   = {Kos Island},
}

@inproceedings{geng2024unveiling,
  title={Unveiling the potential of llm-based ASR on chinese open-source datasets},
  author={Geng, Xuelong and Xu, Tianyi and Wei, Kun and Mu, Bingshen and Xue, Hongfei and Wang, He and Li, Yangze and Guo, Pengcheng and Dai, Yuhang and Li, Longhao and others},
  booktitle={2024 IEEE 14th International Symposium on Chinese Spoken Language Processing (ISCSLP)},
  pages={26--30},
  year={2024},
  organization={IEEE}
}

@inproceedings{fathullah2024prompting,
  title={Prompting large language models with speech recognition abilities},
  author={Fathullah, Yassir and Wu, Chunyang and Lakomkin, Egor and Jia, Junteng and Shangguan, Yuan and Li, Ke and Guo, Jinxi and Xiong, Wenhan and Mahadeokar, Jay and Kalinli, Ozlem and others},
  booktitle={ICASSP 2024-2024 IEEE International Conference on Acoustics, Speech and Signal Processing (ICASSP)},
  pages={13351--13355},
  year={2024},
  organization={IEEE}
}

@article{chu2023qwen,
  title={Qwen-audio: Advancing universal audio understanding via unified large-scale audio-language models},
  author={Chu, Yunfei and Xu, Jin and Zhou, Xiaohuan and Yang, Qian and Zhang, Shiliang and Yan, Zhijie and Zhou, Chang and Zhou, Jingren},
  journal={arXiv preprint arXiv:2311.07919},
  year={2023}
}

@article{chu2024qwen2,
  title={Qwen2-audio technical report},
  author={Chu, Yunfei and Xu, Jin and Yang, Qian and Wei, Haojie and Wei, Xipin and Guo, Zhifang and Leng, Yichong and Lv, Yuanjun and He, Jinzheng and Lin, Junyang and others},
  journal={arXiv preprint arXiv:2407.10759},
  year={2024}
}

@article{bai2024seed,
  title={Seed-asr: Understanding diverse speech and contexts with llm-based speech recognition},
  author={Bai, Ye and Chen, Jingping and Chen, Jitong and Chen, Wei and Chen, Zhuo and Ding, Chuang and Dong, Linhao and Dong, Qianqian and Du, Yujiao and Gao, Kepan and others},
  journal={arXiv preprint arXiv:2407.04675},
  year={2024}
}

@article{xue2024ideal,
  title={Ideal-LLM: Integrating Dual Encoders and Language-Adapted LLM for Multilingual Speech-to-Text},
  author={Xue, Hongfei and Ren, Wei and Geng, Xuelong and Wei, Kun and Li, Longhao and Shao, Qijie and Yang, Linju and Diao, Kai and Xie, Lei},
  journal={arXiv preprint arXiv:2409.11214},
  year={2024}
}

@article{mu2024hdmole,
  title={Hdmole: Mixture of lora experts with hierarchical routing and dynamic thresholds for fine-tuning llm-based asr models},
  author={Mu, Bingshen and Wei, Kun and Shao, Qijie and Xu, Yong and Xie, Lei},
  journal={arXiv preprint arXiv:2409.19878},
  year={2024}
}

@article{shazeer2017outrageously,
  title={Outrageously large neural networks: The sparsely-gated mixture-of-experts layer},
  author={Shazeer, Noam and Mirhoseini, Azalia and Maziarz, Krzysztof and Davis, Andy and Le, Quoc and Hinton, Geoffrey and Dean, Jeff},
  journal={arXiv preprint arXiv:1701.06538},
  year={2017}
}

@article{hu2021lora,
  title={Lora: Low-rank adaptation of large language models},
  author={Hu, Edward J and Shen, Yelong and Wallis, Phillip and Allen-Zhu, Zeyuan and Li, Yuanzhi and Wang, Shean and Wang, Lu and Chen, Weizhu},
  journal={arXiv preprint arXiv:2106.09685},
  year={2021}
}

@inproceedings{yu2023master,
  title={Master-asr: achieving multilingual scalability and low-resource adaptation in asr with modular learning},
  author={Yu, Zhongzhi and Zhang, Yang and Qian, Kaizhi and Wan, Cheng and Fu, Yonggan and Zhang, Yongan and Lin, Yingyan Celine},
  booktitle={International Conference on Machine Learning},
  pages={40475--40487},
  year={2023},
  organization={PMLR}
}

@inproceedings{panayotov2015librispeech,
  title={Librispeech: an asr corpus based on public domain audio books},
  author={Panayotov, Vassil and Chen, Guoguo and Povey, Daniel and Khudanpur, Sanjeev},
  booktitle={2015 IEEE international conference on acoustics, speech and signal processing (ICASSP)},
  pages={5206--5210},
  year={2015},
  organization={IEEE}
}

@article{xu2025fireredasr,
  title={FireRedASR: Open-Source Industrial-Grade Mandarin Speech Recognition Models from Encoder-Decoder to LLM Integration},
  author={Xu, Kai-Tuo and Xie, Feng-Long and Tang, Xu and Hu, Yao},
  journal={arXiv preprint arXiv:2501.14350},
  year={2025}
}

@article{pratap2020mls,
  title={Mls: A large-scale multilingual dataset for speech research},
  author={Pratap, Vineel and Xu, Qiantong and Sriram, Anuroop and Synnaeve, Gabriel and Collobert, Ronan},
  journal={arXiv preprint arXiv:2012.03411},
  year={2020}
}

@article{fedus2022review,
  title={A review of sparse expert models in deep learning},
  author={Fedus, William and Dean, Jeff and Zoph, Barret},
  journal={arXiv preprint arXiv:2209.01667},
  year={2022}
}

@article{zoph2022st,
  title={St-moe: Designing stable and transferable sparse expert models},
  author={Zoph, Barret and Bello, Irwan and Kumar, Sameer and Du, Nan and Huang, Yanping and Dean, Jeff and Shazeer, Noam and Fedus, William},
  journal={arXiv preprint arXiv:2202.08906},
  year={2022}
}

@inproceedings{xie2023moec,
  title={Moec: Mixture of expert clusters},
  author={Xie, Yuan and Huang, Shaohan and Chen, Tianyu and Wei, Furu},
  booktitle={Proceedings of the AAAI Conference on Artificial Intelligence},
  volume={37},
  number={11},
  pages={13807--13815},
  year={2023}
}

@article{song2024u2++,
  title={U2++ moe: Scaling 4.7 x parameters with minimal impact on rtf},
  author={Song, Xingchen and Wu, Di and Zhang, Binbin and Zhou, Dinghao and Peng, Zhendong and Dang, Bo and Pan, Fuping and Yang, Chao},
  journal={arXiv preprint arXiv:2404.16407},
  year={2024}
}

@article{abdin2024phi,
  title={Phi-3 technical report: A highly capable language model locally on your phone},
  author={Abdin, Marah and Aneja, Jyoti and Awadalla, Hany and Awadallah, Ahmed and Awan, Ammar Ahmad and Bach, Nguyen and Bahree, Amit and Bakhtiari, Arash and Bao, Jianmin and Behl, Harkirat and others},
  journal={arXiv preprint arXiv:2404.14219},
  year={2024}
}

@article{park2019specaugment,
  title={Specaugment: A simple data augmentation method for automatic speech recognition},
  author={Park, Daniel S and Chan, William and Zhang, Yu and Chiu, Chung-Cheng and Zoph, Barret and Cubuk, Ekin D and Le, Quoc V},
  journal={arXiv preprint arXiv:1904.08779},
  year={2019}
}

@article{vaswani2017attention,
  title={Attention is all you need},
  author={Vaswani, Ashish and Shazeer, Noam and Parmar, Niki and Uszkoreit, Jakob and Jones, Llion and Gomez, Aidan N and Kaiser, {\L}ukasz and Polosukhin, Illia},
  journal={Advances in neural information processing systems},
  volume={30},
  year={2017}
}

@article{kim2024towards,
  title={Towards Efficient Visual-Language Alignment of the Q-Former for Visual Reasoning Tasks},
  author={Kim, Sungkyung and Lee, Adam and Park, Junyoung and Chung, Andrew and Oh, Jusang and Lee, Jay-Yoon},
  journal={arXiv preprint arXiv:2410.09489},
  year={2024}
}

@article{agarap2018deep,
  title={Deep learning using rectified linear units (relu)},
  author={Agarap, Abien Fred},
  journal={arXiv preprint arXiv:1803.08375},
  year={2018}
}

@article{conneau2020unsupervised,
  title={Unsupervised cross-lingual representation learning for speech recognition},
  author={Conneau, Alexis and Baevski, Alexei and Collobert, Ronan and Mohamed, Abdelrahman and Auli, Michael},
  journal={arXiv preprint arXiv:2006.13979},
  year={2020}
}

@article{loshchilov2017decoupled,
  title={Decoupled weight decay regularization},
  author={Loshchilov, Ilya and Hutter, Frank},
  journal={arXiv preprint arXiv:1711.05101},
  year={2017}
}

@article{van2008visualizing,
  title={Visualizing data using t-SNE.},
  author={Van der Maaten, Laurens and Hinton, Geoffrey},
  journal={Journal of machine learning research},
  volume={9},
  number={11},
  year={2008}
}

@article{peng2024survey,
  title={A survey on speech large language models},
  author={Peng, Jing and Wang, Yucheng and Fang, Yangui and Xi, Yu and Li, Xu and Zhang, Xizhuo and Yu, Kai},
  journal={arXiv preprint arXiv:2410.18908},
  year={2024}
}
\end{document}